\documentclass[12pt,preprint]{aastex}

\usepackage{ulem}

\slugcomment{}

\shorttitle{THE ACS SURVEY OF GLOBULAR CLUSTERS. XX}
\shortauthors{Hempel et al.}

\begin{document}

\title{THE ACS SURVEY OF GLOBULAR CLUSTERS\thanks{Based on observations
  with the NASA/ESA {\it{Hubble Space Telescope}}, obtained at the Space
  Telescope Science Institute, which is operated by AURA, Inc., under NASA
  contract NAS 5-26555, under program GO-10775 (PI: A. Sarajedini).}. {\bf{XIII}}. \\ PHOTOMETRIC CALIBRATION IN
  COMPARISON WITH \\ STETSON STANDARDS }

\author{Maren Hempel\altaffilmark{2,}\altaffilmark{3}}
\affil{Instituto de Astrof\'isica, Pontificia Universidad
  Cat\'olica de Chile,  Facultad de Fisica, Vicu\~na Mackenna 4860, 7820436 Macul, Santiago, Chile}
\email{mhempel@astro.puc.cl}

\author{Ata Sarajedini}
\affil{Department of Astronomy, University of Florida, 211 Bryant Space
  Science Center, Gainesville, FL 32611, USA}
\email{ata@astro.ufl.edu}

\author{Jay Anderson}
\affil{Space Telescope Science Institute, 3700 San Martin Drive, Baltimore, MD
21218, USA}
\email{jayander@stsci.edu}

\author{Antonio Aparicio}
\affil{Department of Astrophysics, University  of La Laguna, V\'ia L\'actea
  s/n, E-38200 La Laguna, Spain}
\email{antapaj@iac.es}

\author{Luigi R. Bedin}
\affil{INAF/Osservatorio Astronomico di Padova, Vicolo dell'Osservatorio 5, I-35122 Padova, Italy }
\email{luigi.bedin@oapd.inaf.it}

\author{Brian Chaboyer}
\affil{Department of Physics and Astronomy, Dartmouth College, 6127 Wilder
  Laboratory, Hanover, NH 03755, USA}
\email{Brian.Chaboyer@dartmouth.edu}

\author{Steven R. Majewski}
\affil{Department of Astronomy, University of Virginia, P.O. Box 400325,
  Charlottesville, VA 22904-4325, USA}
\email{srm4n@virginia.edu}

\author{Antonio Mar\'in-Franch}
\affil{Centro de Estudios de F\'isica del Cosmos de Arag\'on, Plaza. San Juan, 1 Planta-2 | 44001 Teruel, Spain}
\email{amarin@cefca.es}

\author{Antonino Milone}
\affil{Australian National University, College of Physical \& Mathematical Sciences, Canberra ACT 0200
Australia}
\email{antonino.milone@anu.edu.au}

\author{Nathaniel E.Q. Paust}
\affil{Whitman College, 345 Boyer Avenue, Walla Walla, WA 99362, USA}
\email{paustne@whitman.edu}

\author{Giampaolo Piotto}
\affil{Dipartimento di Astronomia, University di Padova, vicolo
  dell'Osservatorio 5, I-35122 Padova, Italy}
\email{giampaolo.piotto@unipd.it }

\author{I. Neill Reid}
\affil{Space Telescope Science Institute, 3700 San Martin Drive, Baltimore, MD
21218, USA}
\email{inr@stsci.edu}

\author{Alfred Rosenberg}
\affil{Instituto de Astrof\'isica de Canarias, V\'ia L\'actea s/n, E-38200 La
  Laguna, Spain}
\email{alf@iac.es}

\and
\author{Michael Siegel}
\affil{Department of Astronomy and Astrophysics, Pennsylvania State University, 525 Davey Laboratory, State College, PA 16801, USA}
\email{siegel@swift.psu.edu}

\altaffiltext{2}{Department of Astronomy, University of Florida, 211 Bryant Space
  Science Center, Gainesville, FL 32611, USA}
\altaffiltext{3}{The Milky Way Millennium Nucleus, Av. Vicu\~{n}a Mackenna 4860,
782-0436 Macul, Santiago, Chile}
\begin{abstract}

In this study we compare the photometric data of 34 Milky Way globular
clusters, observed within the ACS Treasury Program (PI: Ata
Sarajedini) with the corresponding ground-based data, provided by the Photometric Standard Field Catalogs of Stetson
(2000, 2005).\footnote{This research used the facilities of the Canadian
  Astronomy Data Centre operated by the National Research Council of
  Canada with the support of the Canadian Space Agency.} We focus on
the transformation between the HST/ACS F606W to $V$-band and F814W to
$I$-band only. The goal is to assess the validity of the
filter transformation equations by Sirianni et al.\,(2005) with
respect to their dependence on metallicity, Horizontal Branch
morphology, mass and integrated $(V-I)$ color of the various globular
clusters. Such a dependence is expected due to the fact that the
transformation equations are based on the observations of only one
globular cluster, i.e., NGC\,2419. Surprisingly, the
correlation between offset and metallicity is found to be weak, with a
low level significance. The correlation between offset and
Horizontal Branch structure, as well as total cluster mass is still weaker. Based on the
available data we do not find the photometric offset to be linked to  
multiple stellar populations, e.g., as found in NGC\,0288, NGC\,1851, and
NGC\,5139. The results of this study show that there are small systematic offsets
between the transformed ACS- and observed ground based photometry, and that these
are only weakly correlated, if at all, with various cluster parameters and their underlying stellar
populations. As a result, investigators wishing to transform globular cluster
photometry from the Sirianni et al.\ (2005) ground-based $V$, $I$ system onto
the Stetson (2000) system simply need to add --0.040 ($\pm$0.012) to the
$V$ magnitudes and --0.047 ($\pm$0.011) to the $I$ magnitudes. This in turn means that the transformed ACS$V-I$ colors match the
ground-based values from Stetson (2000) to within $\sim$0.01 mag. 

\end{abstract}


\section{Introduction}
\label{sintro}

The Advanced Camera for Surveys on board the Hubble Space Telescope
\citep{pavlovsky04} is one of the cornerstones of astronomical
research. In particular, its high spatial resolution combined with its Wide Field
Channel make it exceptionally well suited to investigate dense stellar
systems like globular clusters, not only in our own Milky Way
\citep[e.g.,][]{sarajedini07}, but in extragalactic systems as well
\citep[e.g.,][]{cote04,jordan07b}. The results of these studies are
often matched with independent data obtained with ground based
instruments to extend the field of view or the wavelength range 
of the observations. However, it is important to note
that the ACS filters differ significantly from ground-based filter sets
and that the photometric transformation between them could depend on knowledge of
the underlying stellar population, e.g., the age or metallicity of the object being
observed. In addition, these photometric data,
resolved and integrated, are used to derive the metallicity and other
parameters of stars and stellar populations. Clearly, it is paramount to estimate 
the accuracy of the photometric transformation procedure. \\

The standard ACS to ground-based photometric transformation equations,
as described by \citet{sirianni05}, are based on a comparison of ACS
photometry in several bands for the Stetson standard field data in
NGC\,2419 \citep[e.g.,][]{stetson00,stetson05}, a massive globular
cluster in the Milky Way
\citep[][]{stetson00,stetson05,baumgart09}. The Stetson data offer
ground-based Johnson/Kron-Cousin $U,B,V,R$, and~$I$ -band photometry of
$>$1300 stars. With a metallicity of $[Fe/H]=~-2.14$
\citep[][]{zinn85,suntzeff88} and total luminosity of $M_V=~-9.42$
\citep[][edition 2010, see also
  http://physwww.mcmaster.ca/~harris/mwgc.dat]{harris96}, NGC\,2419 is
not only one of the most metal-poor, but also the fourth brightest
known Milky Way globular cluster. As recently discovered by
\cite{dicriscienzo11}, NGC\,2419 contains {\it``a large and extreme''}
second stellar generation, representing $\sim$ 30 $\%$ of the total
stellar population, and featuring a different He-content.

Applying filter transformation equations to any photometric data set
based on this cluster alone may therefore result in a systematic
offset between the transformed and directly measured photometry,
reducing the high quality of the original ACS data. These differences,
which we will henceforth call the `photometric offset' or short
`offset', are the subject of this study. Here we compare the
transformed $V$- and $I$-band photometry of various Milky Way globular
clusters, derived from ACS F606W and F814W observations, with their
counterparts by \citet{stetson00,stetson05}, i.e. Johnson $V$, and
Cron-Cousin $I$, respectively (which we will be calling $V$ and $I$
throughout this paper). When referring to the ACS based data, we will
use the term `transformed' $V$ and $I$. Our goal is to search for, and
if confirmed, to quantify the correlation between the photometric
offset and various cluster parameters. \\

The outline of this paper is as follows. In Section \ref{sdata} we
briefly describe the ACS and Stetson data set, and, in more
detail, discuss how both photometric catalogs were matched. Section
\ref{sdepend1} focuses on the photometric offsets and their
correlation with the globular cluster parameters, such as integrated
color, metallicity, and Horizontal Branch structure. Given that HB
structure and metallicity are closely linked, we will investigate the
correlation between photometric offset, metallicity and HB structure
separately for the HB stars only. Due to the mounting evidence that
more massive globular clusters tend to harbor multiple stellar
populations \citep[e.g.,][]{piotto09,milone10,roh11}, we include the
total cluster mass in the list of cluster parameters and discuss the
effect of NGC\,5139, the most massive target cluster, which also known to host
multiple stellar populations
\citep[e.g.,][]{lee99,pancino00,bedin04,dacosta09,bellini10}. Given the wide spread of GC
parameters, Section \ref{sspread} deals with the effects this
diversity may have on the standard deviation of the photometric
offset, and therefore the accuracy of the derived correlation
coefficients. We summarize our findings in Section
\ref{ssummary}.\\

\section{Photometric Sample and Matching Procedure}
\label{sdata}

This study is based on the photometric catalogs for 34 Milky Way
globular clusters (MWGCs) observed in the ACS Survey of Galactic
Globular Clusters (PI: Ata Sarajedini, see Sarajedini et
al.\,2007) and the Photometric Standard Field Catalog by
P. Stetson\footnote{http://www4.cadc.hia.nrc.gc.ca/community/STETSON/standards.}. Detailed
information on the data reduction applied to both samples can be found
in \citet[][]{sarajedini07} and \citet[][]{anderson08} for the ACS
survey and \citet[][]{stetson00,stetson05}, respectively.  The purpose
of this study is to compare the converted $V-$ and $I-$band photometry
of the ACS observations with their counterparts from Stetson and
hence particular care has to be taken when matching the photometry of
individual stars, given the high stellar density within the globular
clusters. In this section we will present the details of the matching
procedure. \\ To match the photometric data for each of the clusters
we use the ACS images, the Stetson source catalogs with RA and DEC
coordinates and the {\texttt{IRAF}}\footnote{IRAF is distributed by the National
  Optical Astronomy Observatories, which are operated by the
  Association of Universities for Research in Astronomy, Inc., under
  cooperative agreement with the National Science Foundation.} task
{\texttt{tfinder}}. The latter is used to adjust the plate solution of the
ACS images to correspond with the RA and DEC coordinates given by
Stetson. The required input information includes: the pixel scale of
the ACS images (0$\farcs05$), the observation equinox (J2000.0), and the
world coordinates and pixel coordinates of the reference pixel within
the ACS image.

\begin{deluxetable}{lcccr}
\tabletypesize{\scriptsize}
\tablewidth{0pt}
\tablecaption{Results of the catalog matching procedure.\label{ttfind}}
\tablehead{
\colhead{Cluster}    &    \colhead{Num$_{PS}^{1}$} &   \colhead{rms$_{RA}$[arcsec]}
&   \colhead{ rm$_{DEC}$[arcsec]}  & \colhead{Num$_{match}^{2}$}}
\startdata
E~~~0003   &        50  &    0.0215775   &    0.0205797   &    207   \\
NGC\,0104   &        12  &    0.0094098   &    0.0157275   &    36    \\
NGC\,0288   &        19  &    0.0106115   &    0.0111964   &    108   \\
NGC\,0362   &        23  &    0.0515514   &    0.0383049   &    118   \\   
NGC\,1261   &        14  &    0.0049360   &    0.0084422   &    49    \\
NGC\,1851   &        19  &    0.0216507   &    0.0147235   &    54    \\ 
NGC\,2298   &        17  &    0.0079661   &    0.0105671   &    86    \\
NGC\,2808   &        15  &    0.0133545   &    0.0175066   &    148   \\
NGC\,3201   &        18  &    0.0129494   &    0.0053085   &    66    \\
NGC\,4147   &        38  &    0.0409929   &    0.0286021   &    255   \\ 
NGC\,4590   &        23  &    0.0141201   &    0.0055300   &    44    \\
NGC\,4833   &        24  &    0.0281217   &    0.0208677   &    35    \\
NGC\,5053   &        23  &    0.0163323   &    0.0304663   &    35    \\
NGC\,5139   &        19  &    0.0198540   &    0.0154449   &    49    \\
NGC\,5272   &        27  &    0.0197264   &    0.0111676   &    46    \\
NGC\,5286   &        20  &    0.0294617   &    0.0171257   &    26    \\
NGC\,5466   &        12  &    0.0688927   &    0.0221558   &    9     \\
NGC\,5904   &        46  &    0.0165915   &    0.0132720   &    260   \\
NGC\,5927   &        19  &    0.0170713   &    0.0138404   &    14    \\ 
NGC\,6093   &        6   &    0.0311338   &    0.0022479   &    6     \\
NGC\,6171   &        7   &    0.0111846   &    0.0030558   &    8     \\
NGC\,6205   &        21  &    0.0168024   &    0.0200496   &    105   \\
NGC\,6341   &        40  &    0.0235042   &    0.0145158   &    631   \\
NGC\,6352   &        21  &    0.0183443   &    0.0204583   &    75    \\
NGC\,6362   &        13  &    0.0182156   &    0.0153341   &    18    \\
NGC\,6397   &        25  &    0.0243064   &    0.0083323   &    112   \\
NGC\,6441   &        23  &    0.0255059   &    0.0174784   &    83    \\
NGC\,6541   &        30  &    0.0335179   &    0.0484386   &    77    \\
NGC\,6752   &        21  &    0.0178754   &    0.0121514   &    46    \\
NGC\,6809   &        41  &    0.0297873   &    0.0173643   &    149   \\
NGC\,6838   &        19  &    0.0247012   &    0.0084462   &    25    \\
NGC\,7078   &        44  &    0.0373419   &    0.0167781   &    239   \\
NGC\,7089   &        32  &    0.0404868   &    0.0290307   &    83    \\
NGC\,7099   &        35  &    0.0279548   &    0.0169936   &    50    \\
\enddata
\tablenotetext{1}{Number of stars used to re-calculate the plate solution (PS) of
  the ACS images.}
\tablenotetext{2}{Number of stars with matched ACS and Stetson photometry,
  i.e., offset position $\le$1.5 pixel.}
\end{deluxetable}

The new plate solution is then used in
{\texttt{IRAF/wcsctran}} to convert the RA and DEC values of the Stetson
catalogs into corresponding xy-pixel coordinates in the ACS
images. This requires that a number ($\gg 3$) of stars in the Stetson
catalog are unmistakably identified in the ACS images, and are widely
distributed over the field of view. Table \ref{ttfind} gives the
number of stars in each cluster used to calculate the new plate
solution (column 2), as well as how much the original RA and DEC for
each star deviates from the one derived from the new plate solution
and the xy-pixel position. There are no multiple matches in any of the merged catalogs, only the
best fitting pair of ACS and Stetson detections are included in the
data base. \\

Although the ACS and Stetson globular cluster samples have 43 objects
in common only 34 were suitable for our study. The remaining 9 GCs (NGC\,5024, NGC\,5986, NGC\,6101, NGC\,6121, NGC\,6218, NGC\,6254,
NGC\,6584, NGC\,6656 and NGC\,6723) do not overlap sufficiently in their
field of view to re-calculate the plate solution and hence to derive
their xy-pixel position with the required accuracy. We note that in
the final catalog, containing both ACS and Stetson
photometry, the RA and DEC coordinates are based on the high quality
ACS astrometry. The Stetson world-coordinates are used as a starting
point to obtain ACS pixel coordinates only.\\

To match the ACS and Stetson photometry we use {\texttt{TOPCAT}}
\citep[e.g.,][]{taylor05} and compare the xy-pixel coordinates of each
detected star using the match option `2-D Cartesian'. We consider a star to
match between both catalogs if the separation between the ACS and
Stetson coordinates is not greater than 1.5 ACS
pixel, i.e. 0\farcs075. The limit is based on the accuracy of the RA
and DEC coordinates (see Table \ref{ttfind}, column 3 and 4) obtained
with {\it{tfinder}}, which is less than 0\farcs05 in RA and DEC,
corresponding to approximately 1 ACS pixel in each dimension. The
2D-pixel positions of matched objects should therefore differ by no
more than 1.5 ACS pixels. Limiting the maximum pixel offset in such a
way will inevitably reduce the number of matched stars in the combined
ACS$-$Stetson catalog, but it will also reduce the probability of
mismatches in the more crowded, central regions of an individual
cluster.

For each cluster in our sample we derived the number of stars with 1
or more neighbors within the matching radius (as applied in the
catalog matching), which we consider bonafide candidates for mismatches. The ratio
between those stars and the total number of stars in the original ACS
catalog is a measure for the possible false matches. Obviously, this mismatch
rate depends on the position of a star, or its local stellar
density. Given that the corresponding Stetson stars, due to the lower
spatial resolution of the data, are mostly found in the outer regions
of the ACS FOV we can assume that the number of mismatches in the
combined ACS$-$ Stetson catalogs is much lower. We find that no cluster
has more than 4$\%$ stars with close neighbors. This ratio is higher for NGC\,6441
(4.36$\%$), but as we will describe in Section \ref{sobserv}, this
cluster will be excluded from further analysis. Including the high
accuracy of the transformed pixel coordinates (see Table \ref{ttfind}) we assume that the closet match, as
found by {\texttt{TOPCAT}}, has a low risk of being a mismatch.  \\

The photometric offset, as used in the further analysis, is always
calculated as the difference between the transformed ACS magnitudes
and its Stetson counterpart: $V_{\rm ACS}$-$V_{\rm Stetson}$ and
$I_{\rm ACS}$-$I_{\rm Stetson}$, respectively. \\

\section{Photometric offsets Relative to Cluster Properties}
\label{sdepend1}

Following the matching of the photometric catalogs, we apply one
additional selection criterion to the stellar sample of each globular
cluster. Based on the ACS photometry, we reject all stars that were
saturated on the ACS images. Even though \citep[][]{gil2004} has shown
that 1\% photometry can be achieved for saturated stars on the ACS
chips, we have decided to be conservative and eliminate these stars
from the comparisons presented in this paper.  As shown in Table
\ref{toffsetgc}, this reduces the total number of stars in NGC\,6093
from 6 to only 2. Therefore, we will exclude this cluster from
later analyses due to its low number statistics.  For comparison we
also derived the photometric offset between the two data sets, as well
as the correlation coefficients for the complete matched cataloges,
i.e. without applying any selection criteria. Although the mean offset
for both filters is smaller, its standard deviation increases. Without
going more into detail we find all correlations studied here to be
weaker than for the selected sample, and will focus on the selected
sample for the remainder of this study.

\begin{deluxetable}{lcccccccccr}
\tabletypesize{\scriptsize}
\tablewidth{0pt}
\tablecaption{Mean and median offset between the convertet ACS $V$ and $I$-band
  photometry and the Stetson standard field catalogs.  \label{toffsetgc}}
\tablehead{ 
\colhead{Number} & \colhead{$\Delta{}V_{mean}$} & \colhead{$\sigma_{V}$}
&\colhead{$\Delta{}I_{mean}$} & \colhead{$\sigma_{I}$}
& \colhead{N$_{match}$} & \colhead{[M/H]$_{GC}^{1}$} &  \colhead{Age$^{2}$}  &\colhead{$\Delta{}Age^{2}$} & \colhead{l [deg]$^{3}$} & \colhead{b [deg]$^{3}$} }
\startdata
E\,0003 &  -0.1345 &  0.0293 &  -0.0990 &   0.0254 &    207$^{4}$ &  -0.69 &  1.02 &  0.15 &  292.27 &  -19.02\\
NGC\,0104 &  -0.0252 &  0.1062 &  -0.0548 &   0.1002 &     19 &  -0.64 &  1.05 &  0.09 &  305.90 &  -44.89\\
NGC\,0288 &  -0.0452 &  0.0477 &  -0.0687 &   0.0517 &    100 &  -0.92 &  0.83 &  0.03 &  152.28 &  -89.38\\
NGC\,0362 &  -0.0610 &  0.1633 &  -0.0634 &   0.1724 &    100 &  -0.87 &  0.81 &  0.05 &  301.53 &  -46.25\\
NGC\,1261 &  -0.0290 &  0.1120 &  -0.0462 &   0.0633 &     38 &  -0.86 &  0.79 &  0.05 &  270.54 &  -52.13\\
NGC\,1851 &  -0.0662 &  0.0667 &  -0.0925 &   0.0861 &     52 &  -0.81 &  0.75 &  0.04 &  244.51 &  -35.04\\
NGC\,2298 &  -0.0351 &  0.0607 &  -0.0498 &   0.0473 &     71 &  -1.49 &  0.99 &  0.05 &  245.63 &  -16.01\\
NGC\,2808 &  -0.0641 &  0.0865 &  -0.0894 &   0.0837 &    108 &  -0.89 &  0.85 &  0.02 &  282.19 &  -11.25\\
NGC\,3201 &  -0.0225 &  0.0640 &  -0.0521 &   0.0311 &     36 &  -1.02 &  0.81 &  0.03 &  277.23 &   8.64\\
NGC\,4147 &  -0.0278 &  0.0412 &  -0.0272 &   0.0371 &    232 &  -1.28 &  0.89 &  0.03 &  252.85 &   77.19\\
NGC\,4590 &  -0.0197 &  0.0304 &  -0.0413 &   0.0199 &     26 &  -1.78 &  0.91 &  0.04 &  299.63 &   36.05\\
NGC\,4833 &  -0.0568 &  0.0722 &  -0.0815 &   0.0701 &     33 &  -1.49 &  0.98 &  0.05 &  303.61 &   -8.01\\
NGC\,5053 &  -0.0285 &  0.0615 &  -0.0477 &   0.0203 &     27 &  -1.76 &  0.96 &  0.04 &  335.69 &   78.94\\
NGC\,5139 &  -0.0321 &  0.0941 &  -0.0721 &   0.1137 &     19 &  -1.13 &  0.89 &  0.05 &  309.10 &   14.97\\
NGC\,5272 &  -0.0438 &  0.0427 &  -0.0586 &   0.0451 &     27 &  -1.12 &  0.89 &  0.04 &   42.21 &   78.71\\
NGC\,5286 &  -0.0183 &  0.0598 &  -0.0506 &   0.0469 &     21 &  -1.19 &  0.98 &  0.05 &  311.61 &   10.57\\
NGC\,5466 &  -0.0450 &  0.0263 &  -0.0460 &   0.0301 &      7 &  -1.98 &  1.07 &  0.05 &   42.15 &   73.59\\
NGC\,5904 &  -0.0727 &  0.0825 &  -0.0815 &   0.0655 &    208 &  -0.90 &  0.83 &  0.02 &    3.86 &   46.80\\
NGC\,5927 &  -0.0584 &  0.0689 &  -0.0547 &   0.1161 &      9 &  -0.50 &  1.01 &  0.11 &  326.60 &    4.86\\
NGC\,6093 &   0.0650 &  0.1273 &   0.0140 &   0.1061 &      2 &  -1.25 &  0.98 &  0.05 &  352.67 &   19.46\\
NGC\,6171 &  -0.0364 &  0.0346 &  -0.0397 &   0.0235 &      8 &  -0.81 &  1.13 &  0.08 &    3.37 &   23.01\\
NGC\,6205 &  -0.0152 &  0.0665 &  -0.0411 &   0.0727 &     51 &  -1.11 &  0.90 &  0.04 &   59.01 &   40.91\\
NGC\,6341 &  -0.0191 &  0.0446 &  -0.0342 &   0.0438 &    564 &  -1.94 &  1.03 &  0.05 &   68.34 &   34.86\\
NGC\,6352 &  -0.0434 &  0.0394 &  -0.0582 &   0.0340 &     61 &  -0.56 &  1.02 &  0.10 &  341.42 &   -7.17\\
NGC\,6362 &  -0.0028 &  0.0276 &   0.0080 &   0.0257 &     14 &  -0.85 &  1.07 &  0.06 &  325.55 &  -17.57\\
NGC\,6397 &  -0.0395 &  0.0244 &  -0.0373 &   0.0248 &    112 &  -1.54 &  0.99 &  0.04 &  338.17 &  -11.96\\
NGC\,6441 &  -0.0803 &  0.0631 &   0.1311 &   0.0842 &     49 &  -0.46 &  0.85 &  0.09 &  353.53 &   -5.01\\
NGC\,6541 &  -0.0678 &  0.0671 &  -0.0791 &   0.0670 &     77 &  -1.31 &  1.01 &  0.04 &  349.48 &  -11.09\\
NGC\,6752 &  -0.0460 &  0.0452 &  -0.0451 &   0.0466 &     36 &  -1.02 &  0.92 &  0.04 &  336.49 &  -25.63\\
NGC\,6809 &  -0.0600 &  0.0525 &  -0.0861 &   0.0619 &    149 &  -1.32 &  0.96 &  0.05 &    8.80 &  -23.27\\
NGC\,6838 &  -0.0254 &  0.0298 &  -0.0275 &   0.0390 &     22 &  -0.59 &  1.11 &  0.10 &   56.74 &   -4.56\\
NGC\,7078 &  -0.0050 &  0.0891 &  -0.0159 &   0.1108 &    123 &  -1.80 &  1.01 &  0.04 &   65.01 &  -27.31\\
NGC\,7089 &  -0.0424 &  0.0769 &  -0.0671 &   0.0767 &     60 &  -1.09 &  0.91 &  0.05 &   53.38 &  -35.78\\
NGC\,7099 &  -0.0415 &  0.0339 &  -0.0477 &   0.0301 &     49 &  -1.70 &  1.01 &  0.04 &   27.18 &  -46.83\\

\bf{Average}  & \bf{-0.0396} & \bf{0.0629} & \bf{-0.0471} & \bf{0.0610} & & & & & &
\enddata
\tablenotetext{1}{From \citet{carretta97}.}
\tablenotetext{2}{From \citet{marin09}.}
\tablenotetext{3}{From \citet{harris96}.}
\tablenotetext{4}{One star was removed from the sample, its offset $>0.5$mag,
  positioned at the edge of the ACS field of view.}
\end{deluxetable}

\subsection{Observational Properties: Magnitude, Color \& Photometric Errors}
\label{sobserv}

\begin{figure*}[h!]
\centering
\includegraphics[scale=.75]{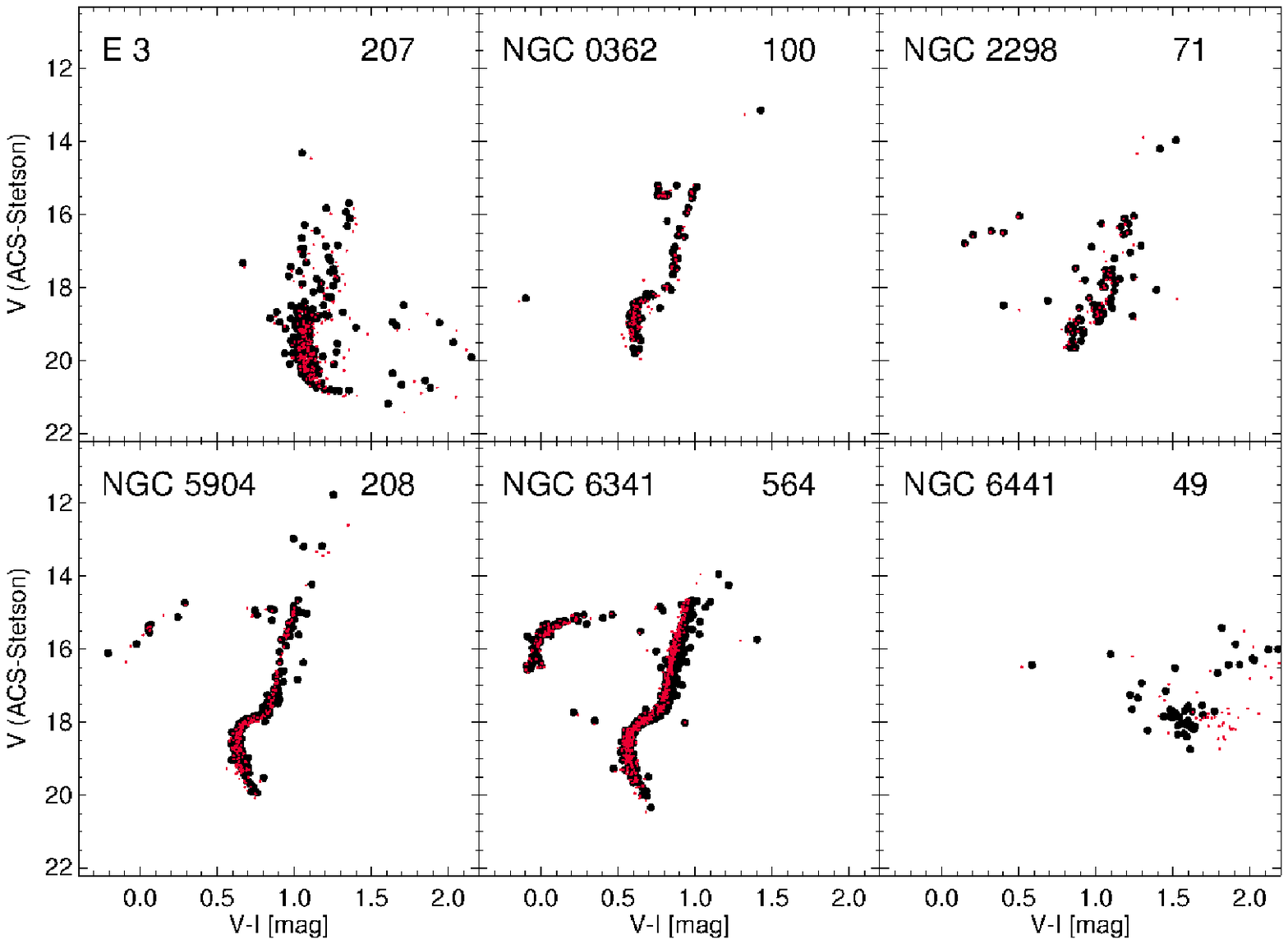}
\caption{$V$~vs.~$(V-I)$ color-magnitude diagrams of 6 selected MWGCs from our
  sample, based on the transformed ACS $V-$ and $I-$band magnitudes (black
  symbols) and their Stetson counterparts (red symbols).   
\label{fcmd1}}
\end{figure*}

\begin{figure*}[ht!]
\centering
\includegraphics[scale=.75]{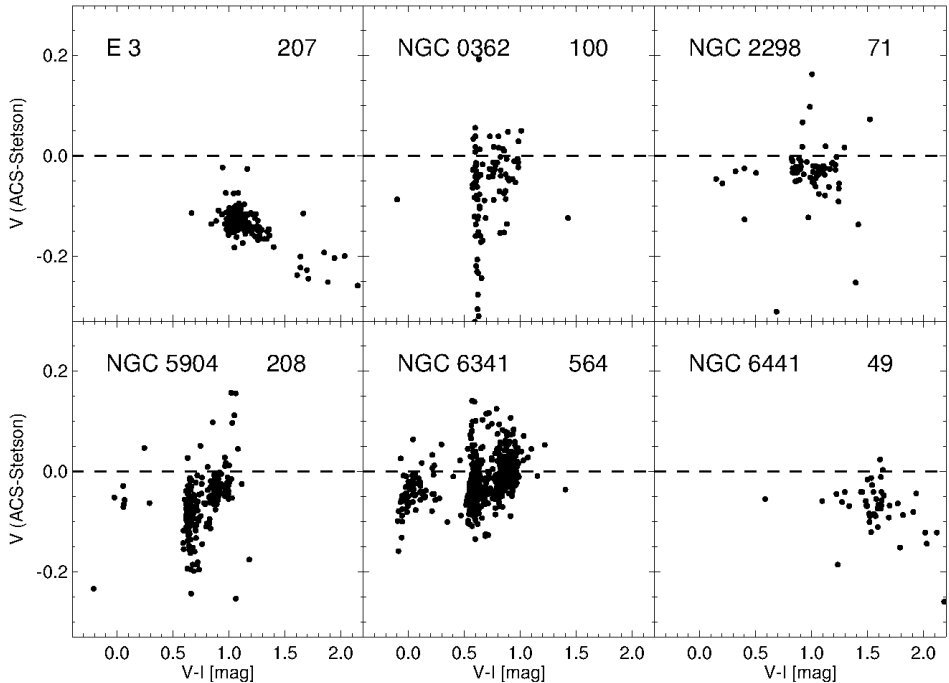}
\caption{V-band magnitude difference (ACS$-$ Stetson) for 6 MWGCs as a
function of the transformed $(V-I)$ color index. The number of stars found in both
data sets is given in the upper right corner of each panel. The horizontal
(dashed)  line marks zero offset.
\label{fdiffV}}
\end{figure*}

\begin{figure*}[ht!]
\centering
\includegraphics[scale=.75]{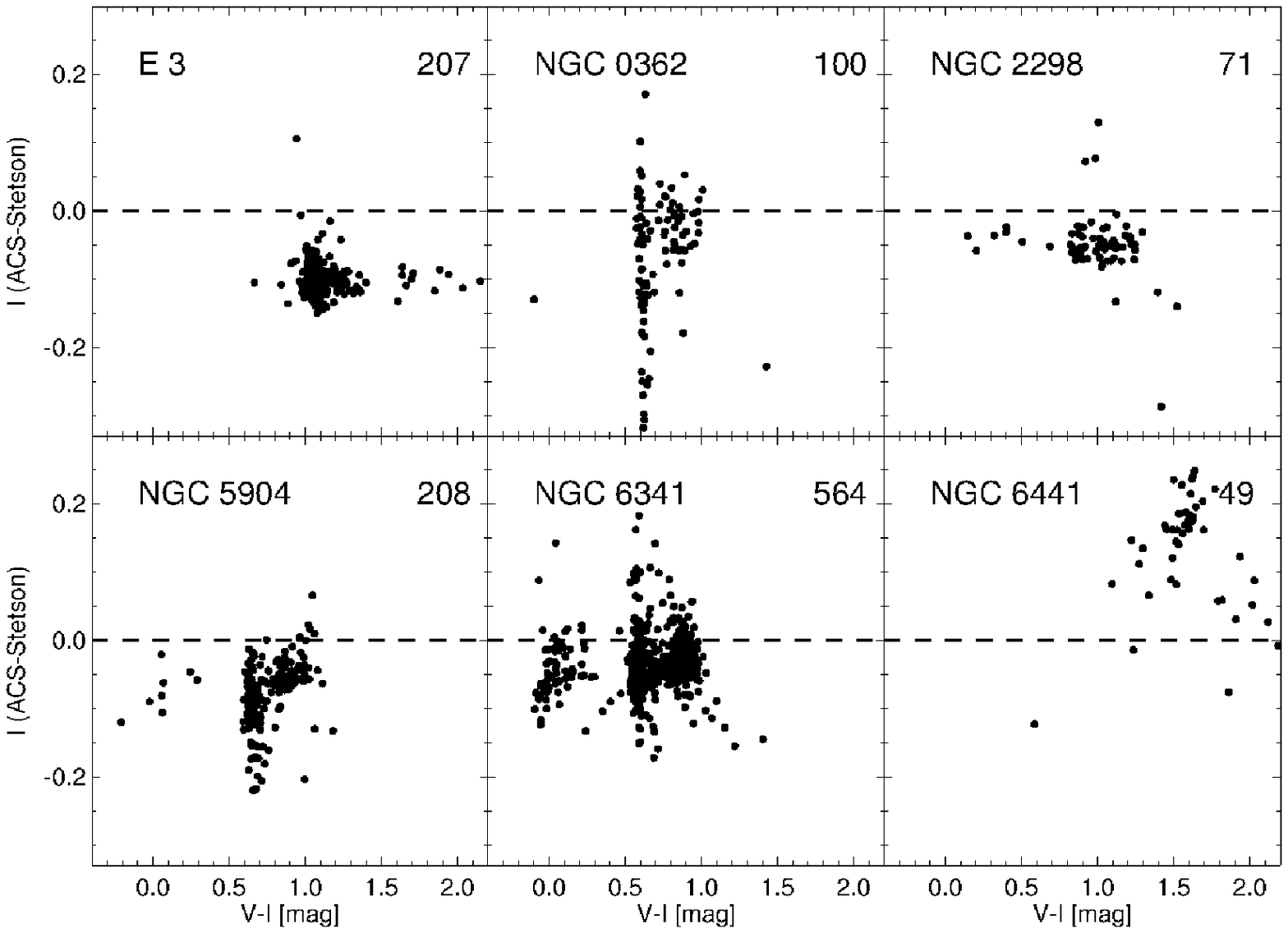}
\caption{Same as Figure \ref{fdiffV}, but for the $I$-band magnituden difference.
\label{fdiffI}}
\end{figure*}

A  comparison between the GC color-magnitude diagrams for selected Globular Clusters\footnote{All plots and matched ACS-Stetson catalogs are available on request.} is shown in
Figure \ref{fcmd1}, where black symbols represent the transformed ACS $V-$
and $I-$ band data, while red symbols correspond to their Stetson
counterparts. In general we find that some
clusters show very good agreement between the space-based and
ground-based photometry (e.g., NGC\,4147, NGC\,5904, NGC\,6397), as well
as some for which either the $V-$ or $I-$ band photometry show a
significant offset (e.g., E\,3), reaching up to $\sim$ 0.3~mag. The
specific offsets become easier to follow when we plot star-by-star
differences between the transformed and observed magnitudes in both
filters, shown for the selected clusters in Figures \ref{fdiffV} and \ref{fdiffI}. For
example, in the case of E\,3, the matched  photometric catalog contains
207 stars following the selection criteria as described in Section \ref{sdata}.

The difference between the transformed ACS $V$-band and the Stetson
photometry shows a clear negative trend with $(V-I)$ color, whereas no
such trend is seen in the $I$-band diagram. On the other hand,
NGC\,6441 features a negative slope in the $\Delta$$V$ vs. $(V-I)$
relation and a positive one in $\Delta$$I$ vs. $(V-I)$. The
transformed $V$-band photometry is also brighter than its Stetson
counterpart, with the difference increasing as color increases. In
contrast, the ACS-$I$- band is fainter, the effect being stronger for
redder stars. This trend differs from that of all of the other
clusters, for which the ACS $I$-band follows the trend of the $V-$
band being the brighter one. At this point, we are unable to trace the
reason for this unusual behavior, but note that the original source
catalog by Stetson is based on the fewest number of observations. We
therefore exclude NGC\,6441 from the correlation analysis.\\

\begin{figure*}[ht!]
\centering
\includegraphics[scale=0.65]{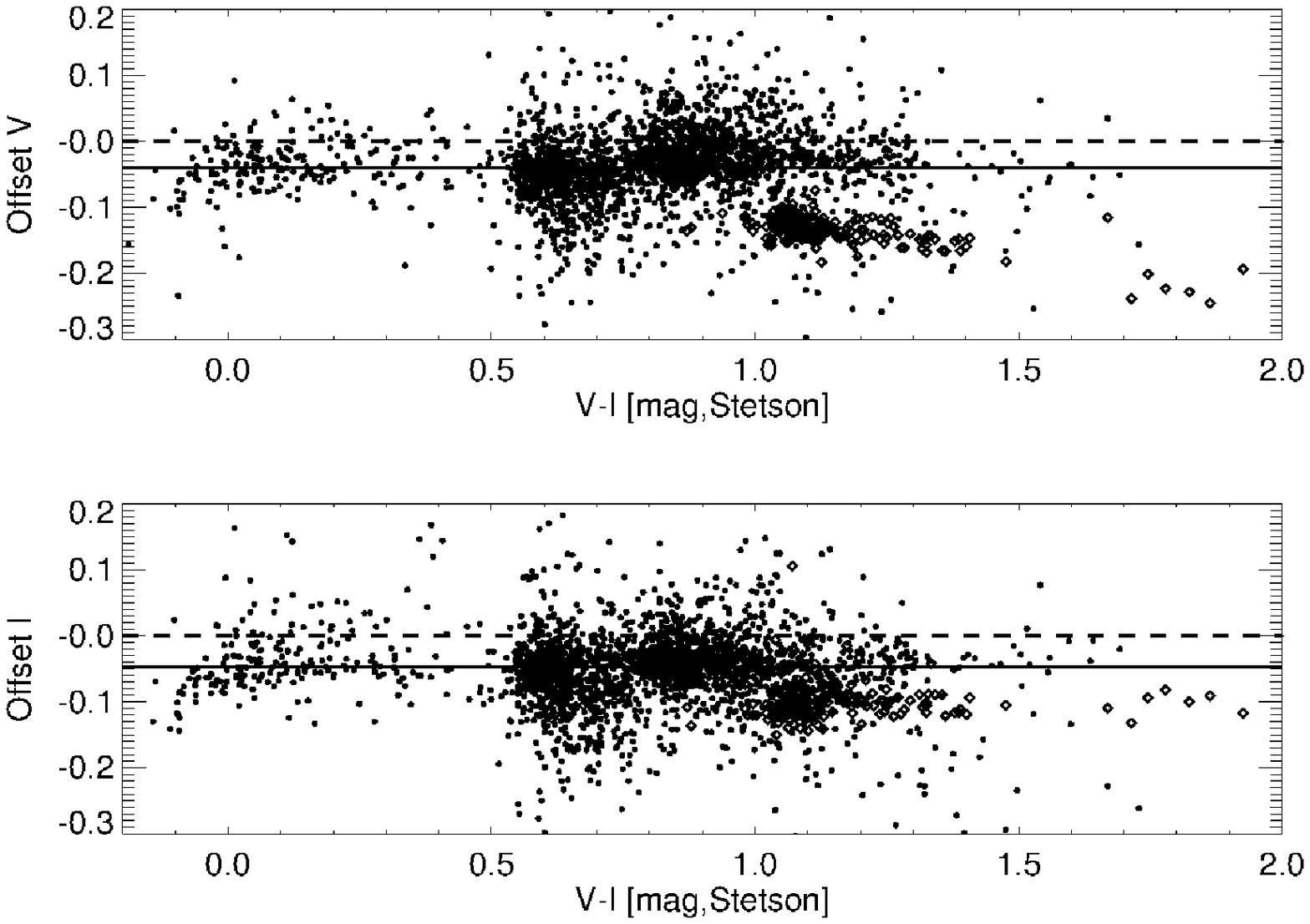}
\caption{Magnitude offsets between the transformed ACS $V-$and $I-$band photometry and the
  Stetson values as a function of the Stetson $(V-I)$ color.  The dashed line marks zero offset, whereas the solid line represents the mean offset derived as the average offset for 34  MWGCs (see Table \ref{toffsetgc}). Open symbols mark stars in the E~3 cluster.
\label{fdiffall}}
\end{figure*}

We combine the data for all of the clusters and compare the
transformed ACS magnitudes with their Stetson counterparts. Figure
\ref{fdiffall} shows the complete sample with the transformed ACS
$(V-I)$ color and the photometric offset in both filters. For both
filters we find a population of stars that do not follow the general
trend, i.e. showing a larger photometric offset than the bulk of the stars
with the same color. These stars are plotted as open diamonds and
belong almost exclusively to E~3. The resulting correlation
coefficients between the offset and the $(V-I)$ color (as given by
Stetson) of the individual stars (excluding E~3), are 0.02 for the
$\Delta$$V$$~vs.$$(V-I)_{\rm Stetson}$ relation and -0.06 for the
$\Delta$$I$$~vs.$$(V-I)_{\rm Stetson}$ relation.

In general, the transformed $I$-band magnitudes are brighter than
their Stetson counterparts, and show a smaller spread around the
`zero' line as compared with the $V$-band. The mean offset for the
complete stellar sample (in total 2461 stars) and their standard deviations (again excluding E~3)
are calculated as 
($\overline{\Delta V}=-0.0382$; $\sigma$=0.0703) and
($\overline{\Delta I}=-0.0519$; $\sigma$=0.0711), which is in agreement with the result based on the average offset for 34 MWGCs.

\begin{figure*}[ht!]
\centering
\includegraphics[scale=.75]{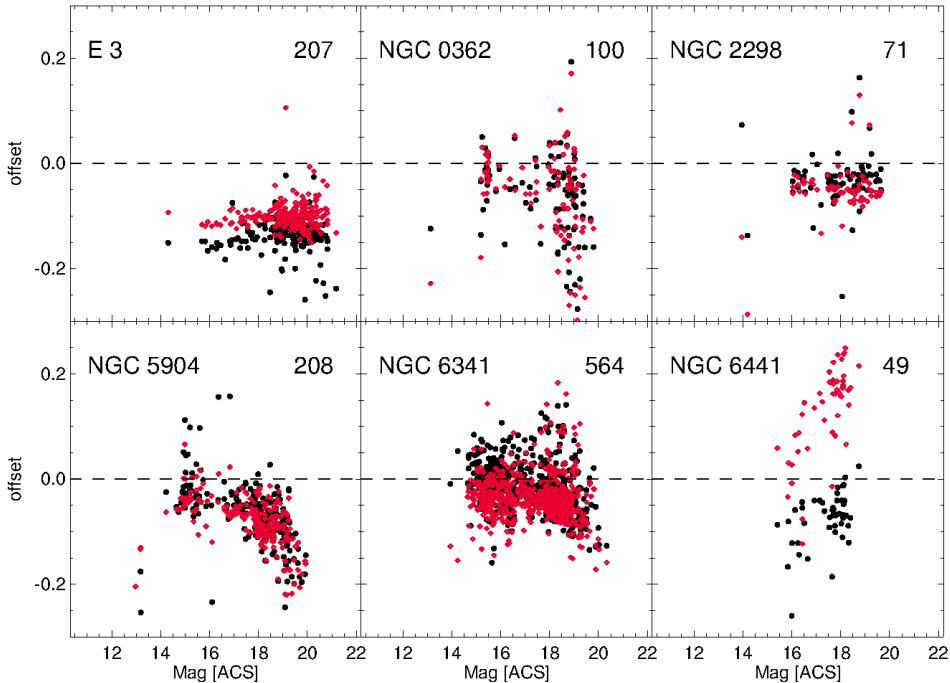}
\caption{Photometric offset between the transformed ACS $V-$and $I-$band and the
  Stetson values as a function of the ACS luminosity. Filled (black) dots refer to the
  $V-$ band , whereas open (red) symbols correspond to the $I-$band. The horizontal
(dashed)  line marks zero offset.
\label{fdiffmag}}
\end{figure*}

\begin{figure*}[ht!]
\centering
\includegraphics[scale=.75]{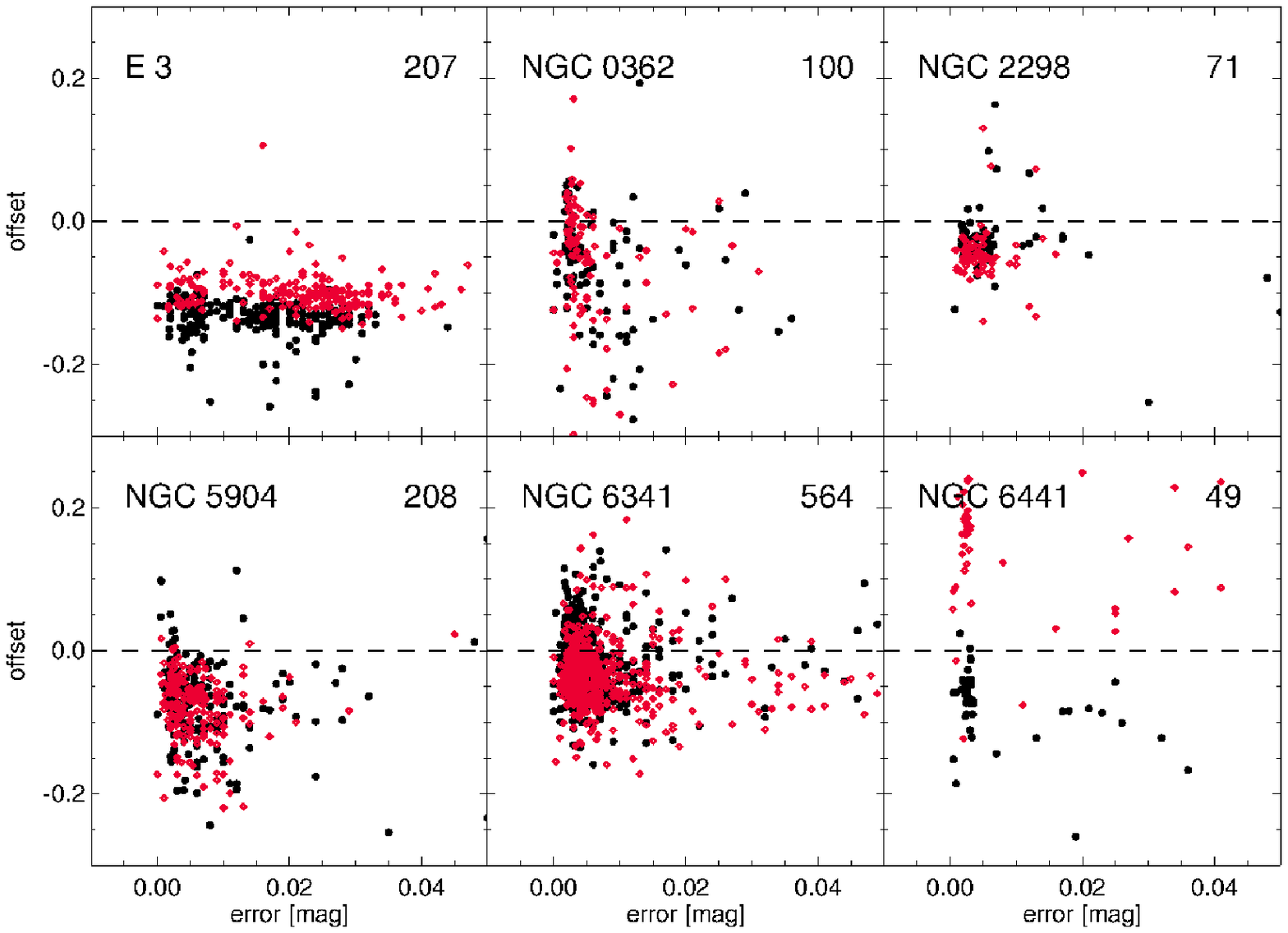}
\caption{Photometric offset in the $V-$and $I-$band as a function of the ACS
  photometric error. The symbols are as in Figure \ref{fdiffmag}.  The horizontal (dashed)  line marks zero offset.
\label{fdifferr}}
\end{figure*}

Before we discuss the correlation between the photometric offsets and
globular cluster properties, we need to consider the possibility that
differences in the photometry are due to the magnitude of the stars
and/or the photometric accuracy. Figures \ref{fdiffmag} and
\ref{fdifferr} show, for each cluster separately, the photometric
offset in both filters as a function of the ACS magnitudes and of the
corresponding photometric errors. As in the $(V-I)$ color plots, we
find a wide range of features. Some, e.g. NGC\,0104, NGC\,2808 show
the largest offset for the brighter, although not saturated stars, and
despite the fact that they have the smallest photometric errors. In
contrast, we find that the brighter stars in NGC\,5904 and NGC\,6809
show smaller photometric offsets as compared with the fainter
ones. However, in general there is no significant systematic
correlation between the photometric errors and the photometric
offsets. \\

\subsection{Integrated and Resolved Color, Metallicity, Mass, Composite Populations}
\label{sparameters}
Our sample of MWGCs allows us to search for correlations between the
photometric offsets and globular cluster properties such as
metallicity, integrated color, horizontal branch morphology, and
globular cluster mass. For our analysis, we use the mean difference
between the transformed ACS magnitude and its counterpart as given by
\citet{stetson00,stetson05}. In addition, we also investigate the
correlations between different GC parameters and the spread of the
offset, i.e., the {\it{standard deviation}}, in particular to assess the
significance of the correlation between the offset and globular
cluster parameters.  In all of the following figures, open symbols
represent the (mean) photometric offset, whereas filled symbols
represent the standard deviation~$\sigma$. A large open triangle marks
NGC\,2419, whose photometry was originally used to derive the standard transformation
equations.  In the discussion of our results we will also include
$Prob_{N}(|r| \geq |r_{0}|)$, the probability (non-directional) that the
Pearson correlation coefficient $|r|$ between the photometric offset
and a given cluster parameter, could also be based on an uncorrelated sample. Hence,
the higher the probability, which we will call the significance level,
the less significant is the correlation. We note that $N$ is the
number of data points ($\gtrsim$~31), and depends on the cluster
parameter, e.g., E~3 does NOT have an integrated $(V-I)$ value.
All of the correlations and related quantities for the $V$-band are listed 
in Table \ref{corrlinV} and those for the $I$-band are in Table \ref{corrlinI}. \\

\begin{deluxetable}{lccr}
\tabletypesize{\scriptsize}
\tablewidth{0pt}
\tablecaption{Linear correlation coefficients between various cluster
  parameters and the photometric offset (mean) and its standard
  deviation in the $V-$band (excluding NGC\,6441 and NGC\,6093, due to its low number of stars).\label{corrlinV}}
\tablehead{
\colhead{Parameter} & \colhead{$\Delta{}V_{mean}$} & \colhead{Prob} & \colhead{\it{stddev$_{V}$}}}
\startdata

$(V-I)$                 & -0.119  & 0.255 & 0.0284    \\
$[M/H]$               & -0.3038 & 0.091 & 0.2243   \\
HBR                   & -0.0234 & 0.451 & -0.2767  \\
$\Delta$$(V-I)$       & -0.0668 & 0.363 & -0.1949  \\
Mass                  & -0.0229 & 0.903 & 0.4700  \\
\enddata
\end{deluxetable}

\begin{deluxetable}{lcccr}
\tabletypesize{\scriptsize}
\tablewidth{0pt}
\tablecaption{Linear correlation coefficients between various cluster
  parameters, the mean photometric offset and its standard
  deviation in the $I-$band (excluding NGC\,6441, NGC\,6093).\label{corrlinI}}
\tablehead{
\colhead{Parameter} & \colhead{$\Delta{}I_{mean}$} & \colhead{Prob} & \colhead{\it{stddev$_{I}$}}}
\startdata

$(V-I)$                 & -0.0905 & 0.299 & -0.0543  \\
$[M/H]$               & -0.2391 & 0.187 & 0.2651   \\
HBR                   & -0.0769 & 0.343 & -0.2748  \\
$\Delta$$(V-I)$       & -0.1748 & 0.178 & -0.2024 \\
Mass                  & -0.3089 & 0.091 & 0.5351  \\
\enddata
\end{deluxetable}

\subsubsection{Integrated / Resolved $(V-I)$ Color}
\label{scolor}
\begin{figure*}[ht!]
\centering
\includegraphics[scale=0.75]{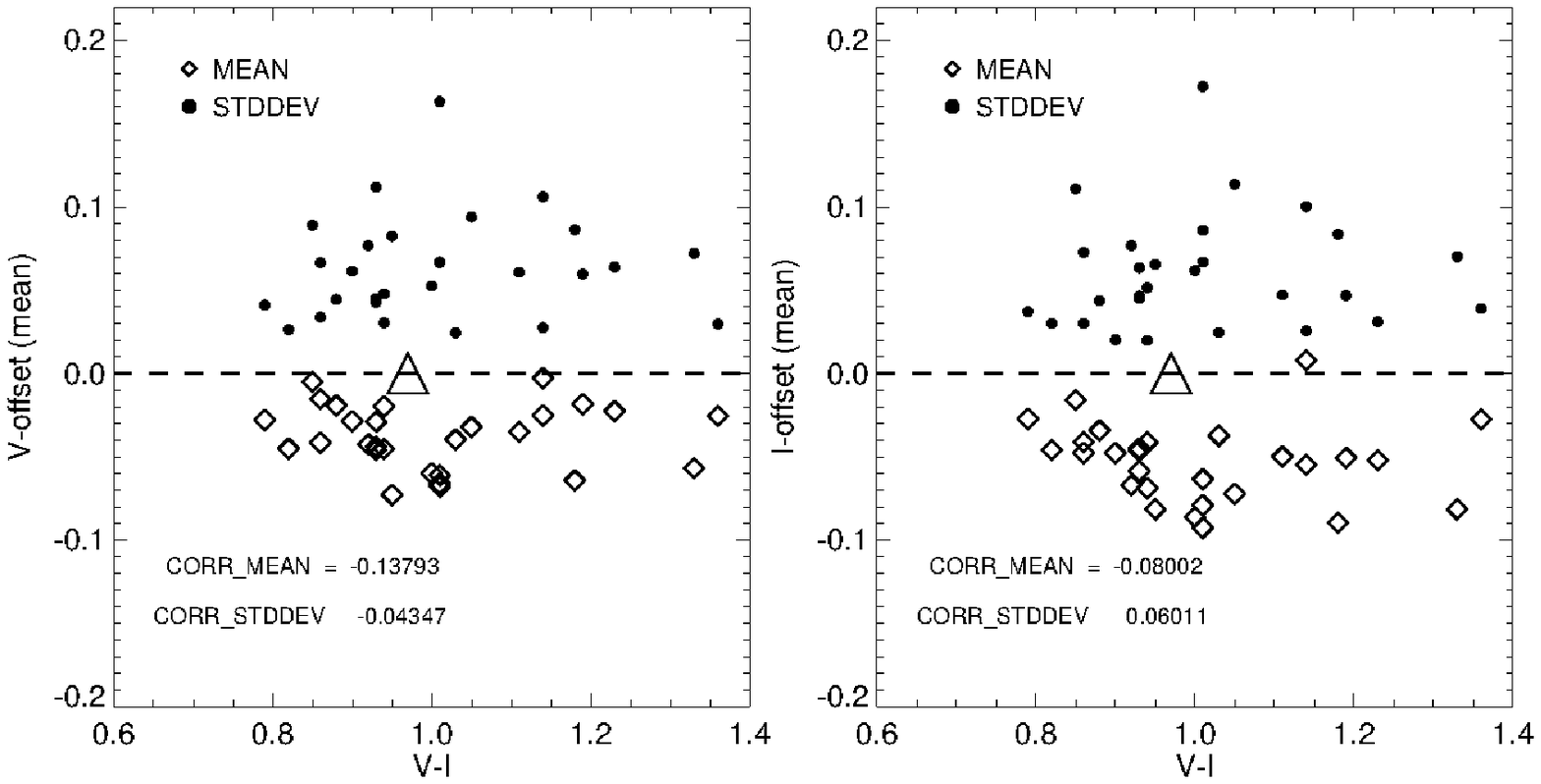}
\caption{Photometric offset in the $V-$ and $I-$band (offset= ACS$-$~Stetson) as a
  function of the globular cluster $(V-I)$ color index. 
  The dashed line represents 'zero'-offset and 
  the open triangle represents NGC\,2419. As before the $(V-I)$ colors
  are corrected for reddening.
\label{foffset_VI}}
\end{figure*}

In Figure \ref{foffset_VI} we show the $\Delta{}V$ and $\Delta{}I$
offsets as a function of the integrated $(V-I)$ color of each cluster
and the corresponding correlation coefficients. We are interested in
this correlation since the integrated color is based only on
observations and does not include any stellar population models, but
also because it depends on the cluster metallicity. We would therefore
expect the correlations between photometric offset and either
integrated color and metallicity to be conform. The $(V-I)$ colors
were corrected for galactic extinction, where both, $(V-I)_{GC}$ and
$E(B-V)_{GC}$ were taken from \cite[][on-line edition
  2010]{harris96}. The latter was converted into $E(V-I)$ using the
prescription by \citet{cardelli89} and \citet{barmby00}:
$E(V-I)$=1.26*$E(B-V)$. \\ The linear correlation coefficient between
the photometric offset and the integrated $(V-I)$ color, as well as
the corresponding significance levels are given in Table
\ref{corrlinV} and Table \ref{corrlinI}.
\vskip 0.5cm

There is no significant correlation between the integrated color and
the photometric offset in our sample, the same is valid for
uncorrected $(V-I)$ colors. However, based on the offset in the
$I$-band (see Figure \ref{foffset_VI}, right panel) clusters bluer
than NGC\,2419 show a much stronger correlation compared to the
complete $V-I$ color range. Selecting only clusters with $(V-I)$
$\le$~$(V-I)$$_{NGC~2419}$, the correlation coefficients are
$corr=-0.3954$; $Prob$=~0.0277 for the $V$-band and $corr=-0.6135$;
$Prob$=~0.0002 for the $I$-band.

In comparison, there is no significant correlation between magnitude offset and
integrated $(V-I)$ cluster color for objects redder then NGC\,2419.\\

\begin{figure*}[ht!]
\centering
\includegraphics[scale=0.5]{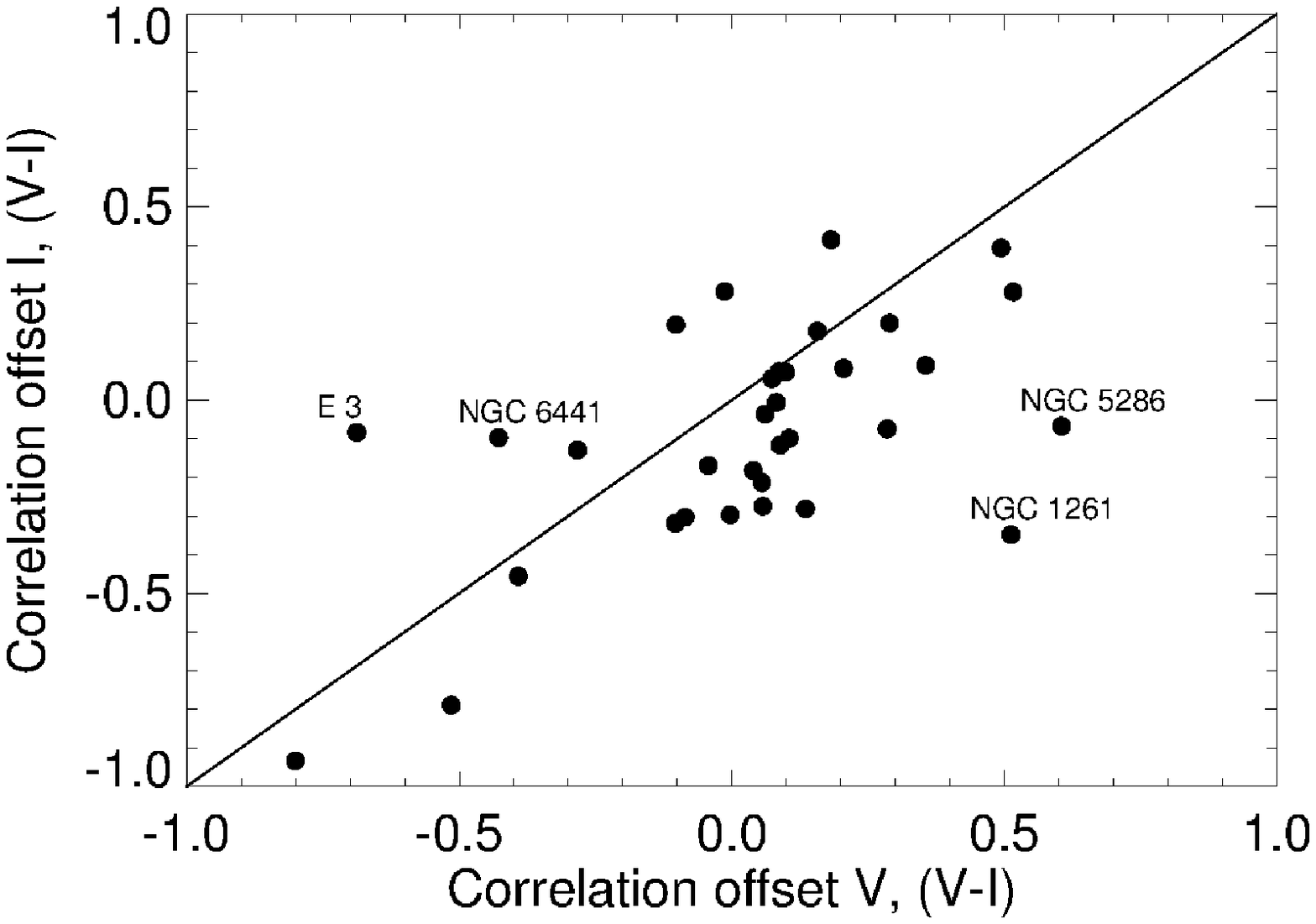}
\caption{Correlation coefficient between $V$ and $I$ band offset and
  the resolved $(V-I)$-color. For E\,3, NGC\,6441, NGC\,1261 and
  NGC\,5286 the correlation between the integrated $V-I$ color and the
  photometric offset in $V$ is not mirrowed by an equally strong
  correlation between $V-I$ and the offset in the $I$-band. Not shown
  is NGC\,6093, due to the low number of contributing stars (2).
\label{corrV_corrI}}
\end{figure*}

Since the integrated globular cluster color depends on the underlying
stellar populations, and also on a proper correction for galactic
extinction, the weak correlations between integrated color and
photometric offset become understandable.  In contrast, when we
compare the correlation coefficients for both the $V$- and $I$- band
filters within an individual cluster, as shown in Figure
\ref{corrV_corrI}, we find them to be consistent. That is to say, a
correlation between $V$-band offset and resolved individual (stellar)
$(V-I)$ color is mirrored by a correlation between $I$-band offset and
color. The correlation between the two filters was derived to be
0.6138, with a high level of significance ($Prob_{N}(|r| \geq
|r_{0}|)$=$~0.00015$).

\begin{figure*}[ht!]
\centering
\includegraphics[scale=0.7]{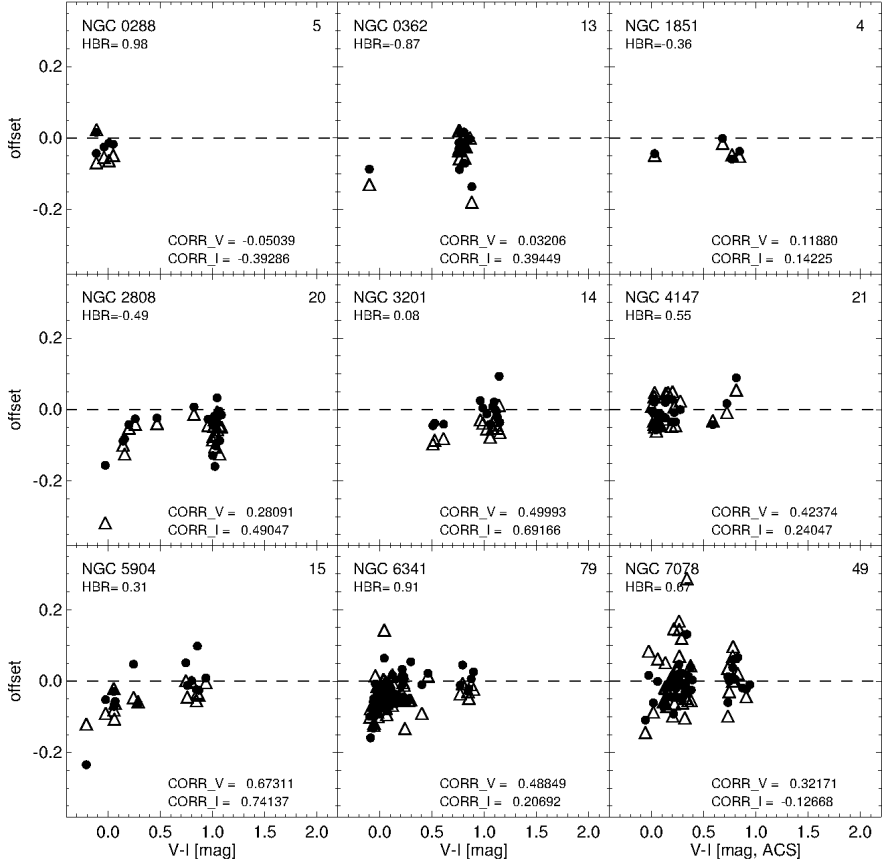}
\caption{Photometric offset in the $V-and~I-$bands for globular clusters with
  a distinct HB. Filled circles refer to the $V$-band, whereas open triangles
  represent the $I$-band data. At the top left we give the name of the cluster
  and its HBR \citep[][]{harris96}. The number of HB stars is given at the
  right. Each panel also gives the linear correlation coefficient between
  $V-I$ color and photometric offset.
\label{hbrsel}}
\end{figure*}

In Figure \ref{hbrsel} we select only the HB stars, because they cover
a wide range of $(V-I)$ color, and also belong to the same
evolutionary phase. We select only clusters with a well defined
Horizontal Branch, e.g., NGC\,0288, NGC\,0362, NGC\,1851, NGC\,2808,
NGC\,3201, NGC\,4147, NGC\,5904, NGC\,6341, and NGC\,7078. We find
that the correlation between the photometric offset and the $(V-I)$
color of the individual stars varies significantly. Interestingly, the
strongest correlation between photometric offset and $(V-I)$ is found
in clusters with the most evenly populated Horizontal Branches, i.e.,
HBR$\sim$0. For example, for NGC\,5904 (HBR=0.31) and NGC\,3201
(HBR=~0.05), the correlation coefficients were estimated to be
[$corr_V=0.673$,~$corr_I=0.741$], and
[$corr_V=0.599$~,~$corr_I=0.692$], respectively. In contrast, clusters
with a very blue or very red Horizontal Branch, e.g., NGC\,0288, or
NGC\,0362, show very little correlation between the photometric offset
and the $(V-I)$ color. We note that these are also the globular
clusters for which the HB structure implies a second parameter spread,
given their very different HBs despite their similar metallicities
(see Section \ref{smetallicity}.)\\

\begin{figure*}
\centering
\includegraphics[scale=0.75]{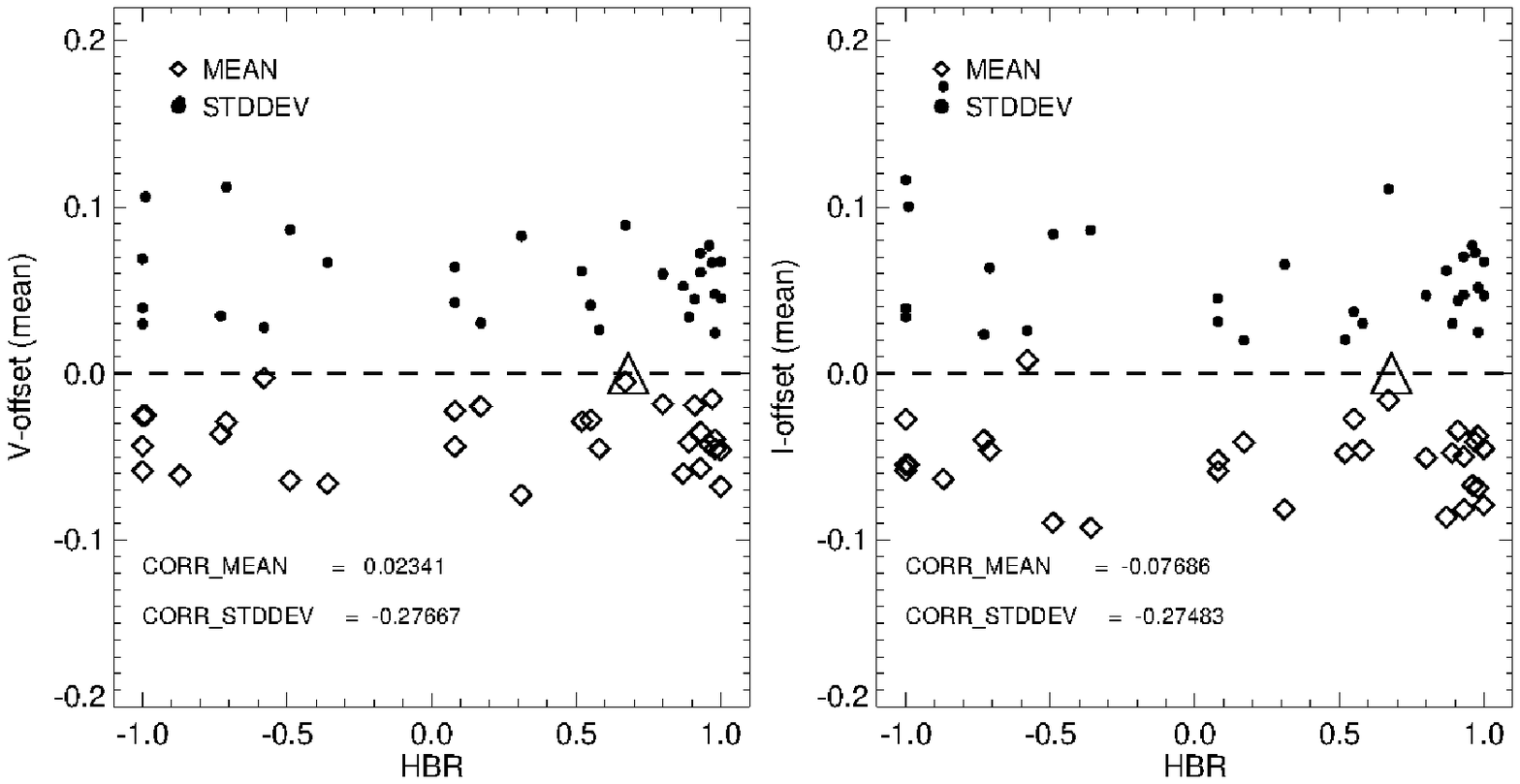}
\caption{Photometric offset in the $V-and~I-$bands as a function of the
  Horizontal Branch Ratio \citep[on-line edition 2003]{harris96}.
\label{foffset_hbr}}
\end{figure*}

\begin{figure*}
\centering
\includegraphics[scale=0.75]{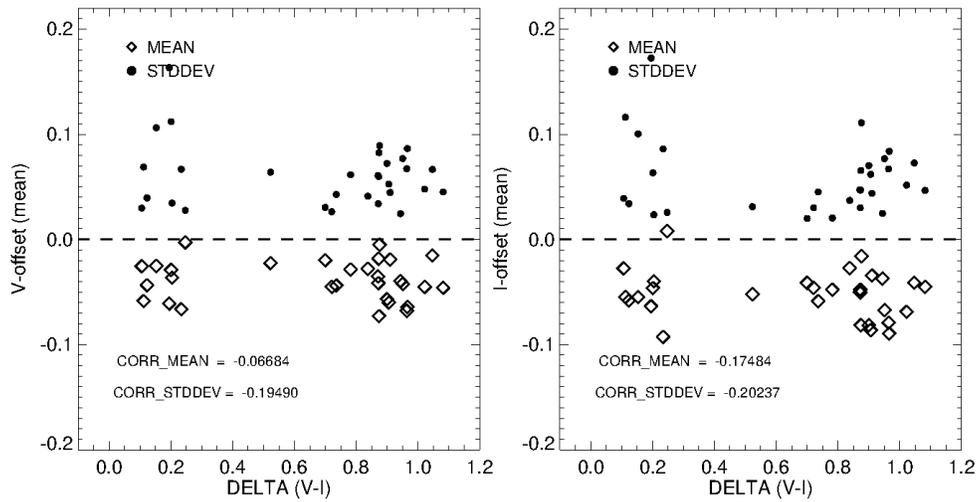}
\caption{Photometric offset in the $V-and~I-$bands as a function of the
  median color difference between HB and the red giant branch \citep[RGB,][]{dotter10}.
\label{foffset_dvi}}
\end{figure*}

The Horizontal Branch morphology is one of the most discerning
globular cluster parameters, i.e. revealing differences in the
chemical composition beyond the matellicity, e.g. in NGC\,0288 and
NGC\,0362 (see above). Therefore, we also examined the correlation
between the photometric offset for each cluster and the Horizontal
Branch structure. To do so we use the Horizontal Branch Ratio
calculated as $HBR=(B-R)/(B+V+R)$ from Harris (1996) (Figure
\ref{foffset_hbr}). In this formulation, $B$ and $R$ denote the
numbers of stars on the $Blue$ or $Red$ side of the RR Lyrae gap,
whereas $V$ represents the number of $Variables$ on the Horizontal
Branch \citep{zinn86,lee90}. Although the quantification of the
Horizontal Branch structure via the HBR is a valuable parameter, in
the case of very blue (e.g., NGC\,0288, NGC\,6341) or very red (e.g.,
NGC\,0362) Horizontal Branches, the HBR can become insensitive to the
HB morphology \citep[][and references therein]{catelan01}. Therefore
we repeat the correlation test and in Figure \ref{foffset_dvi} show
the median $(V-I)$ color difference between the HB and the RGB
$\Delta(V-I)$ \citep[][]{dotter10}.  The correlation coefficients in
both cases are in the range of 0.02 and 0.17, and can therefore be
considered to be negligible.

\subsubsection{Metallicity}
\label{smetallicity}
\begin{figure*}[ht!]
\centering
\includegraphics[scale=0.75]{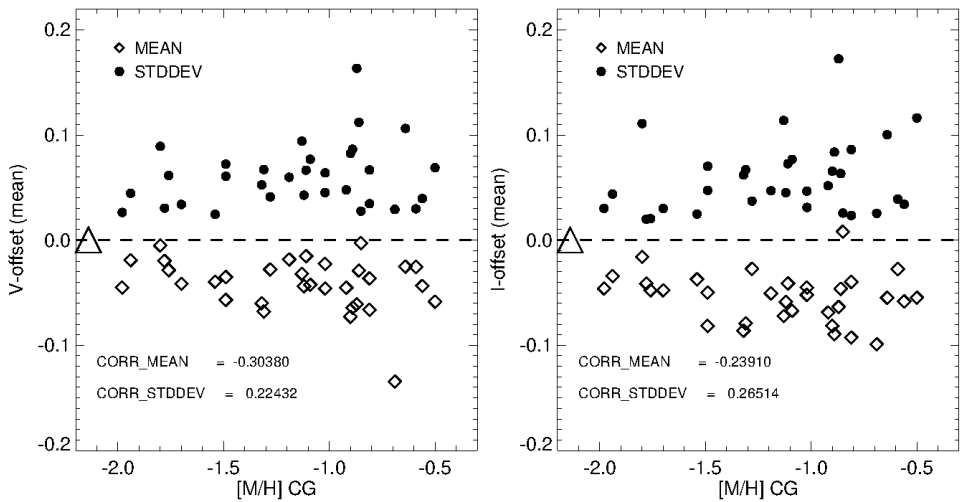}
\caption{Photometric offset in $V-$ and $I-$band (offset= ACS$-$Stetson) as
a function of the globular cluster metallicity. All symbols and colors are
  as in Figure \ref{foffset_VI}. The cluster with the largest offset in the
  $V$-band, at $[M/H]=-0.7$, is E\,3.
\label{foffset_no6441}}
\end{figure*}

As stated by \citet{sirianni05}, transforming accurately between ACS
filters and ground-based filters can be complicated, with
potential dependencies on the stellar spectrum, metallicity, and other
stellar parameters. The metallicity is likely to be important, but
not the only factor playing a role in the filter transformation
equations.  It is also one of the main parameters affecting the
integrated cluster color (see Section \ref{scolor}), as well as the
HB structure (see previous section). The clusters in
our sample have metallicities ranging from $[M/H]=-1.98$ (NGC\,5466) up
to $[M/H]=-0.50$ (NGC\,5927), excluding NGC\,6441 with $[Fe/H]=-0.46$.  In
Figure \ref{foffset_no6441} we show the difference in the two filter
offsets, $V-$ and $I-$, as a function of the metallicity for each
globular cluster. The global metallicities [M/H] were taken from
\citet{marin09} and calculated from [Fe/H] iron abundances using the
prescription by \citet{salaris93}. Based on the correlation coefficients
as given in Table \ref{corrlinV} and Table \ref{corrlinI}, we conclude
that there is a small if any correlation between the $V$- and $I$-band
photometric offsets and the cluster metal abundance. 

\subsubsection{Mass}
\label{smass}

\begin{figure*}[h!]
\centering
\includegraphics[scale=0.75]{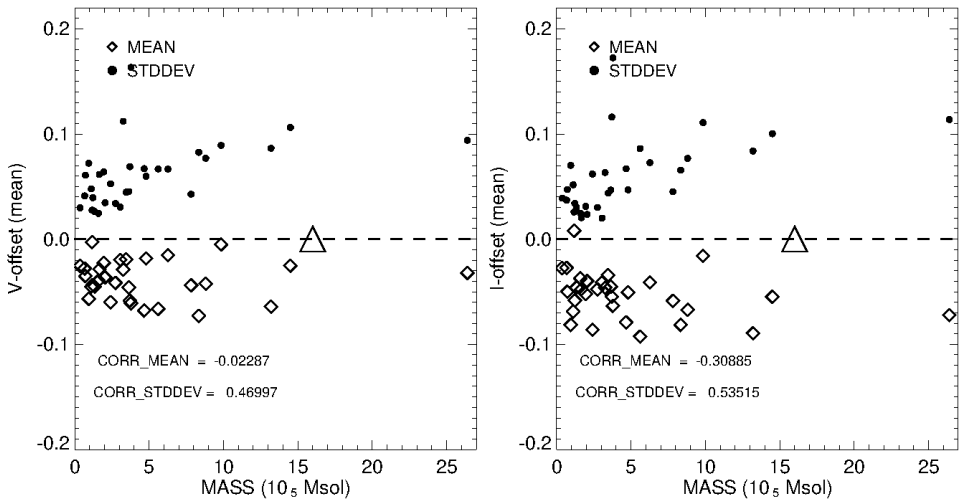}
\caption{Photometric offset in the $V-and~I-$bands as a function of the
 globular cluster mass \citep{gnedin97}.
\label{foffsetmass}}
\end{figure*}

Recent studies have shown that galactic globular clusters are not
single stellar populations in the strictest sense, a fact first
established for the most massive globular clusters. Although two of
the more prominent cases, NGC\,1851 and NGC\,5139, are also part of
this study, it should be noted that NGC\,2419 is also among the more
massive clusters in the Milky Way and has indeed been found to host a
second generation of He-enriched stars \citep[][]{dicriscienzo11}. We
also note that the value for the NGC\,2419 mass, given by
\cite{gnedin97}, 1.6$\times$10$^6$M$_\odot$, differs significantly (up
to $\sim$50$\%$) from the values published by Br\"uns \& Kroupa
(2011), and references therein, which range from 0.9 to 1.19$\times$
10$^6$M$_\odot$ therefore placing the target clusters in relative
context to NGC\,2419 is difficult. However, in Figure
\ref{foffsetmass} we show the photometric offset in the two filter
bands as a function of the total cluster mass \citep[][and references
  therein]{gnedin97}.

The correlation coefficients suggest that there is no statistically
significant correlation between the photometric offsets and the masses
of the globular clusters. There is a slight tendency for more massive
clusters to exhibit a greater dispersion in their magnitude offsets
compared with lower mass clusters; however, this could also be a
result of the possibility that the photometry of more massive clusters
is more likely to be subject to the effects of crowding than that of
less massive clusters.

\section{Extreme Cases of Photometric Offsets}
\label{sspread}

\begin{figure*}[ht!]
\centering
\includegraphics[scale=0.6]{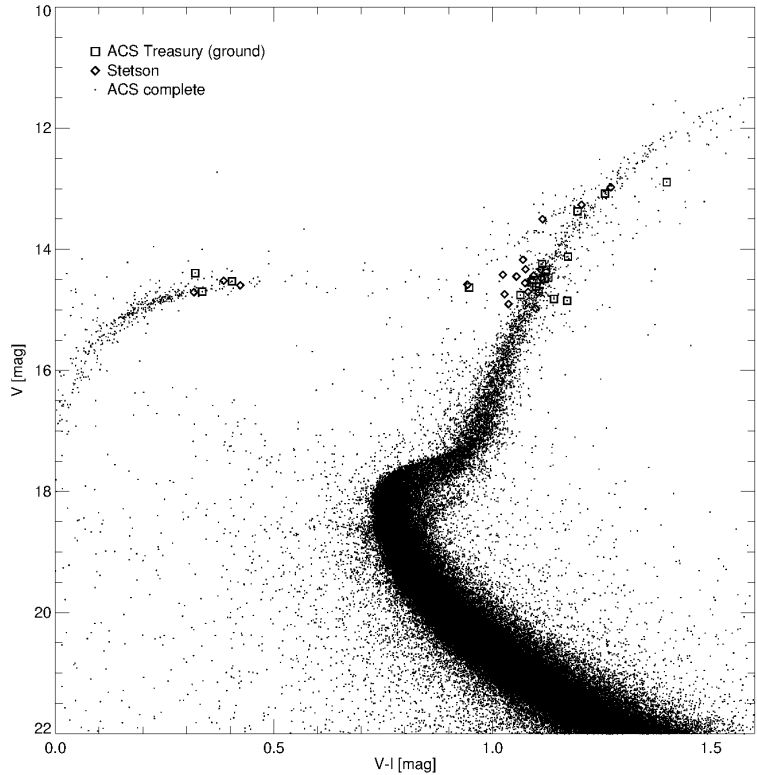}
\caption{$V$vs.$(V-I)$ CMD for NGC\,5139, showing the converted ACS
  photometry as open squares, the Stetson data as open diamonds, and
  the complete ACS sample as dots. Out of 49 stars with matched ACS
  and Stetson photometry only three could tentatively (if at all) be
  assigned to a different RGB then the bulk of the RGB stars.  This is
  likely not sufficient to have any effect on the correlation between
  metallicity and offset.
\label{fn5139}}
\end{figure*}

Our photometric sample is very diverse. Some clusters contain only RGB
stars (e.g., NGC\,0104, NGC\,5139, NGC\,6093), whereas others include
MS, SGB, RGB as well as HB stars (e.g., NGC\,0288, NGC\,5904,
NGC\,6341). The extreme cases, showing the largest discrepancy between
the observed and transformed photometry, or the largest spread
$\sigma$ in photometric offset, are hence of special interest. The
largest offset in both the $V-$ and $I-$ bands is found for the
cluster E\,3. Based on the color magnitude diagram (see Figure
\ref{fcmd1}), the E\,3 sample may include field stars and hence not
represent a single metallicity, used in the correlation test,
i.e. although not being cluster stars but assumed to have the E\,3
metallicity, integrated color and HBR. If those stars are indeed
contaminants and not cluster stars including them in the analysis will
affect the mean offset and consequetially the results of the
correlation tests. This may also be true for several other target
clusters, not due to contamination by field stars, but because they do
not follow the 'single stellar population' paradigm as shown by
\citet[][]{piotto09}, and references therein. In our sample these are
NGC\,0104 \citep[][]{anderson09,dicriscenzio10}, NGC\,0288
\citep[][]{piotto07,roh11}, NGC\,1851 \citep[][]{han09}, and NGC\,5139
\citep[e.g.,][see also Figure
  \ref{fn5139}]{lee99,pancino00,bedin04,bellini10}, which all have
been found to host multiple Main Sequence, Sub-Giant Branch and/or Red
Giant Branch populations. However, in our analysis here the respective
stellar samples are too small to show a significant effect, e.g., an
increased spread in the photometric offset.

Figure \ref{fn5139} shows the color magnitude diagram of NGC\,5139. Of
the 19 stars with both ACS and Stetson photometry, only three may belong
to a slightly redder RGB, i.e. be of higher metallicity than the bulk
of the RGB stars. However, only one of those stars shows a photometric
offset of $\sim$-0.3 mag in the $I$-band, which does not affect the
mean offset or its standard deviation significantly.

The largest dispersion in photometric offset (see Table
\ref{toffsetgc}) is found in a different cluster, NGC\,0362 (with a
sample containing 100 stars). This is somewhat unexpected given that
NGC\,0362 has a very well defined color-magnitude diagram, with no
apparent contaminants. Comparing its properties with the other
clusters of our sample, NGC\,0362 is by no means an exceptional
cluster. Nevertheless, NGC\,0362 (in combination with NGC\,0288) has
been the subject of many studies \citep[][and references
  therein]{stetson96}, due to their different HB morphologies despite
similar metallicities and $\alpha$-abundances. The question remains as
to why NGC\,0362 displays a typical photometric zeropoint offset but
an unusually large standard deviation.

With respect to the minimum discrepancy between transformed and
observed $V$- and $I$-band, the situation is less clear. NGC\,6362,
one of the more metal-rich MWGCs ($[M/H]=-0.85$) shows the smallest
photometric offsets in both filters ($V$-band: $-0.0028/0.0090$,
$I$-band: 0.0080/0.0060 for mean/median offset).  However, the
NGC\,6362 sample contains only 14 stars, which makes this a less solid
result compared with the NGC\,6205 sample, containing 51
stars. NGC\,5053 with 27 stars agrees very well in the $I-$band, and
is with respect to the CMD very similar to NGC\,2419, with a populated
RGB and HB in the combined ACS- and Stetson sample.\\

The available data set shows clearly that the filter transformation
based on NGC\,2419 alone indeed introduces a systematic offset between
the transformed $V$- and $I$-band magnitudes and their observed
counterparts. The correlation between the offset and various cluster
parameters causes deviations between the observed and transformed
magnitudes of up to 0.3~mag, depending on the filter band.

\section{Summary}
\label{ssummary}

The transformation equations traditionally used to convert the ACS
F606W and F814W filters of the Wide-Field Camera into ground-based
Johnson-Cousin $V$- and $I$- band magnitudes are based on observations
of NGC\,2419, one of the most metal-poor and most massive globular
clusters in the Milky Way \citep[][and references
  therein]{harris96}. In our study we confirm the existence of
systematic offsets, expected due to the differences in stellar
spectral energy distributions, in the converted ACS photometry and
their 'original' ground based counterparts. The latter can, in extreme
cases (E\,3), reach up to $\gtrsim$0.3~mag. However, there are no
statistically significant correlations between globular cluster
properties, such as metallicity, and the difference between the
transformed and observed magnitudes. To the extent that they exist,
these correlations seem to be greater in the $V$-band than in $I$. The
strongest correlation has been found between the photometric offset
and the integrated cluster color, followed by metallicity and total
mass.

The
correlation between the photometric offset and the total cluster mass
is less clear and varies between the two filters. We note as well,
that the integrated globular cluster colors \citep{harris96}, as well
as their masses \citep[][and references therein]{gnedin97} are a
compilation of different sources, and hence are less uniform.

The transformation equations by \citet[][]{sirianni05} are based on
the observations of 30-60 NGC\,2419 stars, depending on the filter
band. Here we provide a database of combined ACS and ground based
optical photometry ($V$- and $I$-band), which in some cases (e.g.,
NGC\,6341) included several hundreds of stars, populating the whole
color-magnitude diagram. However, as seen in Figure \ref{fcmd1}, not
only do the number of stars vary widely, but so does the relative
coverage of the color magnitude diagram. NGC\,0288, NGC\,5904 and
NGC\,6341 have the largest number of stars with combined ACS and
ground based photometry, as well as the widest range of evolutionary
stages, including MS, SGB, RGB, and HB. NGC\,6093, which includes only
two stars on the RGB, has been excluded from the correlation
analysis. Additional tests have shown that if we restrict our sample
to only stars above the MSTO, and hence mimickin the NGC\,2419 sample
used by \citet[][]{sirianni05} more closely, the various linear
correlations between the photometric offset and the cluster parameters
are unchanged. In a separate series of correlation tests, we excluded
NGC\,5139 from the cluster sample, given that this MWGC is known to
host various stellar populations \citep[see][and references
  therein]{piotto09}, featuring partly different
metallicities. However, the only correlation that is significantly
affected is the one between mass and the photometric offset. This is
not surprising, given that NGC\,5139 is also by far the most massive
cluster in our sample, and as a result defines the correlation at the
high mass end. In all other correlation tests,
rejecting NGC\,5139 changed the results insignificantly. \\

As described in Section \ref{sdepend1} we exclude saturated stars from
the analysis presented here. However, the correlation test for an
unselected sample, i.e. without rejecting saturated stars, finds all
correlations to be weaker, with a smaller mean offset, but a larger
standard deviation for both filters.

The results of this study show that there are small systematic offsets
between transformed ACS and observed ground based photometry, and that
these are only weakly correlated, if at all, with various cluster
parameters and their underlying stellar population. As a result,
investigators wishing to transform globular cluster photometry from
the Sirianni et al.\ (2005) ground-based $V$, $I$ system onto the
Stetson (2000) system simply need to add --0.040 ($\pm$0.012) to the
$V$ magnitudes and --0.047 ($\pm$0.011) to the $I$ magnitudes. The
quoted errors in each case represent the average value of the standard
errors of each mean offset. This in turn means that the transformed
ACS $V-I$ colors match the ground-based values from Stetson (2000) to
within $\sim$0.01 mag. We note that these offsets are the average of
the mean offset for all clusters in our sample (see also Table
\ref{toffsetgc}). In contrast to that the results in Section
\ref{sobserv} are the mean offset for all individual stars, excluding
E\,3 and NGC\,6441. However, within the photometric errors the average
and mean offset are in agreement.

\acknowledgments

Support for this work (proposal GO-10775) was provided by NASA through a grant
from the Space Telescope Science Institute, which is operated by the
Association of Universities for Research in Astronomy, Inc. under NASA
contract NAS5-26555. Support for MH has been provided by the BASAL Center for
Astrophysics and Associated Technologies PFB-06, the FONDAP Center for
Astrophysics N. 15010003, and the Ministry for the Economy, Developement and Tourism's Programa Iniciativa Cient{\'i}fica Milenio through grant P07-021-F, awarded to The Milky Way Millenium Nucleus. SRM was funded by NSF grant AST-0807945.

\end{document}